# Representations of $Spin(4)$, $Spin(2,2)$ and $Spin(3,1)$

**Ali Delbaznasab, and MohammadReza Molaei**

Mahani Mathematical Research Center, Shahid Bahonar University of Kerman

76169-14111 Kerman, Iran

e-mails: delbaznasab@gmail.com, mrmolaei@uk.ac.ir

**Abstract.** In this essay we present an outer product on $\mathbb{R}^6$, and we show that $\mathbb{R}^6$ with this outer product can take three kinds of Lie brackets. We prove that $\mathbb{R}^6$ with these Lie brackets take the structures of $so(4)$, $so(2,2)$ and $so(3,1)$. We also deduce a $4-$dimensional submanifold of $\mathbb{R}^6$ with an almost complex structure.

## 1. Introduction

In the first section of this paper we introduce an outer product on $\mathbb{R}^6$. This kind of outer product has some properties look alike of usual outer product of $\mathbb{R}^3$. We show that $\mathbb{R}^6$ with this outer product forms a Lie algebra that is isomorphic to the Lie algebra $so(4)$ of the Lie group $So(4)$.

Indeed we construct an associative algebra via a new product on $\mathbb{R}^8$, and we show that this associative algebra is isomorphic to the product of quaternion algebra $H \times H$ (or Cliford algebra $cl(1,2) \cong cl(3)$) [1,7].

We obtain two $8 \times 8$ real representation and $4 \times 4$ unitary complex representation of $Spin(4)$.

In section three we deduce three kind of associative algebras and we deduce few representations of the connected components of the identities of the Lie groups $Spin(2,2)$, $Spin(3,1)$ and $Spin(1,3)$.

In section four we present a $4-$dimensional submanifold of $\mathbb{R}^6$ via the outer product which is considered in section two, and we show that this submanifold has an almost complex structure.

## 2. Representation of $Spin(4)$

Let us begin this section by the following definition.

**Definition 2.1.** Let $a = (a_1, a_2, a_3, a_4, a_5, a_6)$, $b = (b_1, b_2, b_3, b_4, b_5, b_6)$ be two vectors in $\mathbb{R}^6$ then we define the outer product of $a$ and $b$ by $a \times b = (x_1, x_2, x_3, x_4, x_5, x_6)$ where
$x_1 = -(a_2 b_4 - a_4 b_2 + a_3 b_5 - a_5 b_3)$,
$x_2 = a_1 b_4 - a_4 b_1 - (a_3 b_6 - a_6 b_3)$,
$x_3 = a_1 b_5 - a_5 b_1 + a_2 b_6 - a_6 b_2$,
$x_4 = -(a_1 b_2 - a_2 b_1 + a_5 b_6 - a_6 b_5)$,
$x_5 = -(a_1 b_3 - a_3 b_1) + a_4 b_6 - a_6 b_4$,
$x_6 = -(a_2 b_3 - a_3 b_2 + a_4 b_5 - a_5 b_4)$.
We also define $\sigma(a) = \bar{a} = (a_6, -a_5, a_4, a_3, -a_2, a_1)$. We denote the usual inner product of $a$ and $b$ by $a.b$. In this case we have following simple theorem.

**Theorem 2.1.** If t is a real number and $a, b, c \in \mathbb{R}^6$ then
1. $a \times b = -b \times a$ and so $a \times a = 0$,
2. $ta \times b = t(a \times b)$,
3. $a \times (b + c) = a \times b + a \times c$,
4. $(a \times b) \cdot c = a \cdot (b \times c)$,
5. $(a \times b) \cdot a = b \cdot (a \times b)$,
6. $\overline{a \times b} = a \times \bar{b} = \bar{a} \times b$,
7. $a \cdot \bar{b} = \bar{a} \cdot b$,
8. $(a \times b) \times c = (c \cdot a)b - (c \cdot b)a + (c \cdot \bar{a})\bar{b} - (c \cdot \bar{b})\bar{a}$,
9. If we define $[a, b] = a \times b$ then $\mathbb{R}^6$ with this bracket is a Lie algebra isomorphic to $so(4)$.

If $a_0, a_1, b_0, b_1 \in \mathbb{R}$ and
$$p = (a_2, a_3, a_4, a_5, a_6, a_7), q = (b_2, b_3, b_4, b_5, b_6, b_7) \in \mathbb{R}^6$$
then we denote $x = (a_0, a_1, \ldots, a_7)$ and $y = (b_0, b_1, \ldots, b_7)$ by $x = a_0 e_0 + a_1 e_1 + p$, $y = b_0 e_0 + b_1 e_1 + q$, respectively, where $e_0 = (1,0,0,0,0,0,0,0)$, $e_1 = (0,1,0,0,0,0,0,0)$ and $p = (a_2, \ldots, a_7) \cong (0,0,a_2, \ldots, a_7)$.
With these notations we have the next definition.

**Definition 2.2.** We define the product of $x$ and $y$ by:
$x * y = (a_0 e_0 + a_1 e_1 + p)(b_0 e_0 + b_1 e_1 + q) = (a_0 b_0 - a_1 b_1 - p \cdot q)e_0 + (a_0 b_1 + a_1 b_0 - p \cdot \bar{q})e_1 + a_0 q + b_0 p + a_1 \bar{q} + b_1 \bar{p} + p \times q$.

**Theorem 2.3.** $(\mathbb{R}^8, +, \cdot, *)$ is an associative algebra with an identity, where its product is the scalar product of $\mathbb{R}^8$.

**Proof.** We only prove the associatively, the proofs of the other properties are simple.

$(x * y) * z = ((a_0 e_0 + a_1 e_1 + p) * (b_0 e_0 + b_1 e_1 + q)) * (c_0 e_0 + c_1 e_1 + r) = [(a_0 b_0 + a_1 b_1 - p \cdot q) e_0 + (a_0 b_1 + a_1 b_0 - p \cdot \bar{q}) e_1 + a \cdot q + b \cdot p + a_1 \bar{q} + b_1 \bar{p} + p \times q] * [c_0 e_0 + c_1 e_1 + r] = [a_0 b_0 c_0 + a_1 b_1 c_0 - (p \cdot q) c_0 + a_0 b_1 c_1 + a_1 b_0 c_1 - (p \cdot \bar{q}) c_1 - a_0 (q \cdot r) - b_0 (p \cdot r) - a_1 (\bar{q} \cdot r) - b_1 (\bar{p} \cdot r) - (p \times q) \cdot r] e_0 + [a_0 b_0 c_1 + a_1 b_1 c_1 - (p \cdot q) c_1 + a_0 b_1 c_0 + a_1 b_0 c_0 - (p \cdot \bar{q}) c_0 - a_0 (q \cdot \bar{r}) - b_0 (p \cdot \bar{r}) - a_1 (\bar{q} \cdot \bar{r}) - b_1 (\bar{p} \cdot \bar{r}) - (p \times q) \cdot \bar{r}] e_1 + (a_0 e_0 + a_1 e_1 - p.q) r + c_0 (a_0 q + b_0 p + a_1 \bar{q} + b_1 \bar{p} + p \times q) + (a_0 b_1 + a_1 b_0 - p \cdot \bar{q}) \bar{r} + c_1 (a_0 \bar{q} + b_0 \bar{p} + a_1 \bar{\bar{q}} + b_1 \bar{\bar{p}} + \overline{p \times q}) + a \cdot (q \times r) + b \cdot (p \times r) + a_1 (\bar{q} \times r) + b_1 (\bar{p} \times r) + (p \times q) \times r$.

By using of the equalities $p \cdot \bar{q} = \bar{p} \cdot \gamma, \bar{p} \cdot \bar{q} = p \cdot q, \bar{\bar{p}} = p, (p \times q) \times \bar{r} = (p \times \bar{q}) . r, (p \times q) \times r - p \times (q \times r) = -(r \times p) \times q$, and so on we have:

$(x * y) * z = -(p \cdot q) r - (p \cdot \bar{q}) \bar{r} + (p \times q) \times r + (r \cdot q) p + (r \cdot \bar{q}) \bar{p} - p \times (q \times r) = -(r \times p) \times q + (q \cdot r) p - (q \cdot p) r + (q \cdot \bar{r}) \bar{p} - (q \cdot \bar{p}) \bar{r} = 0$. ∎

By Definition 2.2 if $x = (a_0, a_1, \ldots, a_7)$ and $y = (b_0, b_1, \ldots, b_7)$ then $x * y = (x_0, x_1, \ldots, x_7)$ where $x_i$ ($0 \leq i \leq 7$) are defined by:

$x_0 = a_0 b_0 + a_1 b_1 - a_2 b_2 - a_3 b_3 - a_4 b_4 - a_5 b_5 - a_6 b_6 - a_7 b_7$,
$x_1 = a_0 b_1 + a_1 b_0 - a_2 b_7 + a_3 b_6 - a_4 b_5 - a_5 b_4 + a_6 b_3 - a_7 b_2$,
$x_2 = a_0 b_2 + a_2 b_0 + a_1 b_7 + a_7 b_1 - (a_3 b_5 - a_5 b_3 + a_4 b_6 - a_6 b_4)$,
$x_3 = a_0 b_3 + a_3 b_0 - a_1 b_6 - a_6 b_1 + a_2 b_5 - a_5 b_2 - (a_4 b_7 - a_7 b_4)$,
$x_4 = a_0 b_4 + a_4 b_0 + a_1 b_5 + a_5 b_1 + a_2 b_6 - a_6 b_2 + a_3 b_7 - a_7 b_3$,
$x_5 = a_0 b_5 + a_5 b_0 + a_1 b_4 + a_4 b_1 - (a_2 b_3 - a_3 b_2 + a_6 b_7 - a_7 b_6)$,
$x_6 = a_0 b_6 + a_6 b_0 - a_1 b_3 - a_3 b_1 - (a_2 b_4 - a_4 b_2) + a_5 b_7 - a_7 b_5$,
$x_7 = a_0 b_7 + a_7 b_0 + a_1 b_2 + a_2 b_1 - (a_3 b_4 - a_4 b_3 + a_5 b_6 - a_6 b_5)$.

Let $A$ be an associative algebra with multiplicative identity $e$ and let $W$ be a complex vector space. A representation of $A$ on $W$ is determind by a set of linear operators $T(a)$ on $W$ such that

1. $T(\gamma a + \mu b) = \gamma T(a) + \mu T(b)$ such that $a, b \in A$ and $\gamma, \mu \in \mathbb{C}$,
2. $T(ab) = T(a) T(b)$,
3. $T(e) = E$ where $E$ is an identity operator of $W$ [2, 6].

For any $a = (a_1, \ldots, a_8) \in \mathbb{R}^8$ we define a right multiplication map $F_a: \mathbb{R}^8 \mapsto \mathbb{R}^8$ by $x \mapsto x * a$. This is a linear map and if $a = a_0 e_0 + a_1 e_1 + p$, then $F_a$ according to the standard basis of $\mathbb{R}^8$ is represented by the following matrix:

$$A = \begin{bmatrix} a_0 & a_1 & a_2 & a_3 & a_4 & a_5 & a_6 & a_7 \\ a_1 & a_0 & a_7 & -a_6 & a_5 & a_4 & -a_3 & a_2 \\ -a_2 & -a_7 & a_0 & a_5 & a_6 & -a_3 & -a_4 & a_1 \\ -a_3 & a_6 & -a_5 & a_0 & a_7 & a_2 & -a_1 & -a_4 \\ -a_4 & -a_5 & -a_6 & -a_7 & a_0 & a_1 & a_2 & a_3 \\ -a_5 & -a_4 & a_3 & -a_2 & a_1 & a_0 & a_7 & -a_6 \\ -a_6 & a_3 & a_4 & -a_1 & -a_2 & -a_7 & a_0 & a_5 \\ -a_7 & -a_2 & a_1 & a_4 & -a_3 & a_6 & -a_5 & a_0 \end{bmatrix}, \quad (2.1)$$

The set of all matrices as in (2.1) is a real representation of the algebra $(\mathbb{R}^8, +, \cdot, *)$. Also the matrix $A$ can be partitioned in four submatrixs of the form $\begin{bmatrix} U & V \\ -V & U \end{bmatrix}$ where

$$U = \begin{bmatrix} a_0 & a_1 & a_2 & a_3 \\ a_1 & a_0 & a_7 & -a_6 \\ -a_2 & -a_7 & a_0 & a_5 \\ -a_3 & a_6 & -a_5 & a_0 \end{bmatrix} \quad \text{and} \quad V = \begin{bmatrix} a_4 & a_5 & a_6 & a_7 \\ a_5 & a_4 & -a_3 & a_2 \\ a_6 & -a_3 & -a_4 & a_1 \\ a_7 & a_2 & -a_1 & -a_4 \end{bmatrix}.$$

Under the mapping $\varphi: A = \begin{bmatrix} U & V \\ -V & U \end{bmatrix} \mapsto U + iV$ we have following complex representation:

$$B = \begin{bmatrix} a_0 + a_4 i & a_1 + a_5 i & a_2 + a_6 i & a_3 + a_7 i \\ a_1 + a_5 i & a_0 + a_4 i & a_7 - a_3 i & -a_6 + a_2 i \\ -a_2 + a_6 i & -a_7 - a_3 i & a_0 - a_4 i & a_5 + a_1 i \\ -a_3 + a_7 i & a_6 + a_2 i & -a_5 - a_1 i & a_0 - a_4 i \end{bmatrix}. \quad (2.2)$$

**Theorem 2.3.** Let

$$M = \{A \mid A \text{ has the form (2.1)}\} \cong \{B \mid B \text{ has the form (2.2)}\}.$$

Then $M \cong H \times H$ where $H$ is a quaternion algebra.

**Proof.** Let $\varphi: (\mathbb{R}^8, +, \cdot, *) \to H \times H$ be defined by $\varphi(a_0, a_1 \ldots, a_7) = ((a_0 - a_1) + (a_2 - a_7)i + (a_6 + a_3)j + (a_4 - a_5)k, (a_0 + a_1) + (a_2 + a_7)i - (a_3 - a_6)j - (a_4 + a_5)k)$. Then one can easily prove that $\varphi$ is an isomorphism. ∎

**Theorem 2.4.** Let

$$G = \{A \mid A = \varphi(x), \det(A) = 1, a_0 a_1 + a_2 a_7 - a_3 a_6 + a_4 a_5 = 0\}.$$

Then $G$ is isomorphic to the Lie group Spin(4).

**Proof.** Let $\psi: (\mathbb{R}^8, +, \cdot, *) \simeq M \to H \times H$ be the isomorphism of the previous theorem and $\psi(a) = (\psi_1(a), \psi_2(a)) \in H \times H$. We prove that $\psi: M \to H \times H$ is an isomorphism. We know that Spin(4) is isomorphic to $Su(2) \times Su(2)$ or $Sp(1) \times Sp(1)$. If

$$\det(A) = 1 \text{ and } a_0 a_1 + a_2 a_7 - a_3 a_6 + a_4 a_5 = 0 \quad . \tag{2.3}$$

Since

$$\det(A) = [(a_0 + a_1)^2 + (a_2 - a_7)^2 + (a_4 - a_5)^2 + (a_6 + a_3)^2]^2$$
$$\times [(a_0 - a_1)^2 + (a_2 + a_7)^2 + (a_4 + a_5)^2 + (a_6 - a_3)^2]^2$$
$$= \left[\sum_{i=0}^{7} a_i^2 - 2(a_0 a_1 + a_2 a_7 + a_4 a_5 - a_3 a_6)\right]^2 \left[\sum_{i=0}^{7} a_i^2 + 2(a_0 a_1 + a_2 a_7 + a_4 a_5 - a_3 a_6)\right]^2,$$

then (2.3) implies that
$1 = \sum_{i=0}^{7} a_i^2 = (a_0 + a_1)^2 + (a_2 - a_7)^2 + (a_4 - a_5)^2 + (a_6 + a_3)^2 = (a_0 - a_1)^2 + (a_2 + a_7)^2 + (a_4 + a_5)^2 + (a_6 - a_3)^2$.
So $|\psi_1(a)|^2 = |\psi_2(a)|^2 = 1$. Thus $\psi(a) \in S^3 \times S^3$. Since $\psi$ is an isomorphism and $G = \psi^{-1}(S^3 \times S^3)$, then $G$ is the Lie group Spin(4). ∎

**3. Representation of the connected components of the identities of $Spin(2,2)$, $Spin(3,1)$ and $Spin(1,3)$**

As the pervious section we define three brackets $[\,,\,]_1$, $[\,,\,]_2$, $[\,,\,]_3$ on $\mathbb{R}^6$ such that $\mathbb{R}^6$ with this brackets will take three Lie algebras structure isomorphic to $So(2,2)$, $So(3,1)$, $So(1,3)$, respectively.

Let $a = (a_0, a_1, \ldots, a_7)$, and $y = (b_0, b_1, \ldots, b_7) \in \mathbb{R}^8$. Set $p = (a_2, \ldots, a_7)$, $q = (b_2, \ldots, b_7)$ and $L_{ij} = a_i b_j - a_j b_i$. Then we define $[p,q]_1 = p \times_1 q$, $[p,q]_2 = p \times_2 q$, $[p,q]_3 = p \times_3 q$ as the following forms:
$[p,q]_1 = (L_{35} + L_{46}, L_{25} - L_{47}, L_{26} + L_{37}, -L_{23} - L_{67}, -L_{24} + L_{57}, L_{34} + L_{56})$,
$[p,q]_2 = (L_{46} - L_{35}, L_{25} + L_{47}, L_{26} + L_{37}, -L_{23} + L_{67}, -L_{24} + L_{57}, -L_{34} - L_{56})$,
$[p,q]_3 = (-L_{35} - L_{46}, L_{25} - L_{47}, L_{26} + L_{37}, L_{23} - L_{67}, L_{24} + L_{57}, L_{34} - L_{56})$.
We also define
$\sigma_1(p) = \bar{p} = (a_7, -a_6, a_5, a_4, -a_3, a_2)$,

$\sigma_2(p) = \bar{p} = (-a_7, a_6, a_5, -a_4, -a_3, a_2),$
$\sigma_3(p) = \bar{p} = (a_7, -a_6, a_5, -a_4, a_3, -a_2),$
$\langle p, q \rangle_1 = a_2 b_7 + a_3 b_6 - a_4 b_5 - a_5 b_4 + a_6 b_3 + a_7 b_2,$
$\langle p, q \rangle_2 = \langle p, q \rangle_3 = a_2 b_7 - a_3 b_6 + a_4 b_5 + a_5 b_4 - a_6 b_3 + a_7 b_2.$

If $a = a_0 e_0 + a_1 e_1 + p$, $b = b_0 e_0 + b_1 e_1 + q$ we have three product of $a, b$ by:

$a *_1 b = (a_0 b_0 + a_1 b_1 - \langle p, q \rangle_1) e_0 + (a_0 b_1 + a_1 b_0 - \langle p, \sigma_1(q) \rangle_1) e_1 + a_0 q + b_0 p + a_1 \sigma_1(q) + b_0 \sigma_1(p) + [p, q]_1$. For $i = 1, 2$ we have

$a *_i b = (a_0 b_0 + a_1 b_1 - \langle p, q \rangle_i) e_0 + (a_0 b_1 + a_1 b_0 - \langle p, \sigma_i(q) \rangle_i) e_1 + a_0 q + b_0 p + a_1 \sigma_i(q) + b_0 \sigma_i(p) + [p, q]_i.$

Thus we have following representations of the above algebras respectively.

$$A_1 = \begin{bmatrix} a_0 & a_1 & a_2 & a_3 & a_4 & a_5 & a_6 & a_7 \\ a_1 & a_0 & a_7 & -a_6 & a_5 & a_4 & -a_3 & a_2 \\ -a_2 & -a_7 & a_0 & a_5 & a_6 & -a_3 & -a_4 & a_1 \\ a_3 & -a_6 & a_5 & a_0 & a_7 & a_2 & -a_1 & a_4 \\ a_4 & a_5 & a_6 & -a_7 & a_0 & a_1 & a_2 & -a_3 \\ a_5 & a_4 & -a_3 & -a_2 & a_1 & a_0 & a_7 & a_6 \\ a_6 & -a_3 & -a_4 & -a_1 & -a_2 & -a_7 & a_0 & -a_5 \\ -a_7 & -a_2 & a_1 & a_4 & -a_3 & a_6 & -a_5 & a_0 \end{bmatrix}, \quad (3.1)$$

$$A_2 = \begin{bmatrix} a_0 & a_1 & a_2 & a_3 & a_4 & a_5 & a_6 & a_7 \\ -a_1 & a_0 & -a_7 & a_6 & a_5 & -a_4 & -a_3 & a_2 \\ -a_2 & -a_7 & a_0 & a_5 & a_6 & -a_3 & -a_4 & a_1 \\ -a_3 & a_6 & -a_5 & a_0 & a_7 & a_2 & -a_1 & -a_4 \\ a_4 & -a_5 & a_6 & a_7 & a_0 & -a_1 & a_2 & a_3 \\ -a_5 & -a_4 & a_3 & -a_2 & a_1 & a_0 & a_7 & -a_6 \\ a_6 & a_3 & -a_4 & a_1 & -a_2 & a_7 & a_0 & a_5 \\ a_7 & -a_2 & -a_1 & -a_4 & -a_3 & -a_6 & -a_5 & a_0 \end{bmatrix}, \quad (3.2)$$

$$A_3 = \begin{bmatrix} a_0 & a_1 & a_2 & a_3 & a_4 & a_5 & a_6 & a_7 \\ -a_1 & a_0 & a_7 & -a_6 & a_5 & -a_4 & a_3 & -a_2 \\ a_2 & -a_7 & a_0 & a_5 & a_6 & a_3 & a_4 & -a_1 \\ a_3 & a_6 & -a_5 & a_0 & a_7 & -a_2 & a_1 & a_4 \\ a_4 & -a_5 & -a_6 & -a_7 & a_0 & -a_1 & -a_2 & -a_3 \\ -a_5 & -a_4 & a_3 & -a_2 & a_1 & a_0 & a_7 & -a_6 \\ -a_6 & a_3 & a_4 & -a_1 & -a_2 & -a_7 & a_0 & a_5 \\ -a_7 & -a_2 & a_1 & a_4 & -a_3 & a_6 & -a_5 & a_0 \end{bmatrix}. \quad (3.3)$$

So we have the following theorem.

**Theorem 3.1.** Let

$$G_1 =$$
$$\{A_1 \mid A_1 \text{ is the same as } (3.1), \det(A_1) = 1, a_0 a_1 + a_2 a_7 + a_3 a_6 - a_4 a_5 = 0\},$$
$$G_2 =$$
$$\{A_2 \mid A_2 \text{ is the same as } (3.2), \det(A_2) = 1, a_0 a_1 + a_2 a_7 + a_3 a_6 + a_4 a_5 = 0\}.$$
Then

1. $G_1$ is isomorphic to the connected component of the identity of the Lie group Spin(2,2).

2. $G_2$ is isomorphic to the connected component of the identity of the Lie group Spin(3,1) $\cong$ Spin(1,3).

**Proof.** One can easily prove this theorem such as Theorem 2.1, but in the case 1 we can define $\varphi: G_1 \to Sl(2, \mathbb{R}) \times Sl(2, \mathbb{R})$ by $\varphi(A_1) = (A, B)$ where
$$A = \begin{bmatrix} a_0 - a_1 + a_3 + a_6 & a_4 - a_5 - a_2 + a_7 \\ a_4 - a_5 + a_2 - a_7 & a_0 - a_1 - a_3 - a_6 \end{bmatrix},$$
$$B = \begin{bmatrix} a_0 + a_1 + a_3 - a_6 & -a_4 - a_5 - a_2 - a_7 \\ -a_4 - a_5 + a_2 + a_7 & a_0 + a_1 - a_3 + a_6 \end{bmatrix},$$
We can easily prove that $\varphi$ is the desired isomorphism. ∎

Spin(3,1) is called the Lorentz group in some books. Similar to above theorem, there is a representation for Spin(3,1).
$$\varphi(a_0, a_1 \ldots, a_7) = ((a_0 - a_1) + (a_2 - a_7)i + (a_4 - a_5)j + (a_6 + a_3)k, (a_0 + a_1) + (a_2 + a_7)i + (a_4 + a_5)j + (a_6 - a_3)k),$$
where $j^2 = k^2 = 1$. We know that [4] Spin(2,2) is isomorphic to $Sl(2, \mathbb{R}) \times Sl(2, \mathbb{R})$.

Such as case 2 we can define $\psi: G_2 \to \text{Sl}(2, \mathbb{C})$ by
$$\psi(A_2) = \begin{bmatrix} a_0 - a_7 + (a_1 + a_2)i & -a_3 - a_4 + (a_5 + a_6)i \\ a_3 - a_4 + (a_5 - a_6)i & a_0 + a_7 + (a_1 - a_2)i \end{bmatrix},$$
where $\text{Sl}(2, \mathbb{C}) = \{A \mid A \in GL(2, \mathbb{C}), \det(A) = 1\}$.

Since the connected component of the identity of Spin(2,2) is isomorphic to $\text{Sl}(2, \mathbb{C})$ [4], then we can easily show that $\psi$ is our desired isomorphism. ∎

## 4. Almost complex structure on a 4-dimensional submanifold of $\mathbb{R}^6$

In this section we define a 4-dimensional submanifold of $\mathbb{R}^6$ which admit an almost complex structure.

Let $M = \{x = (x_1, \ldots, x_6) \in \mathbb{R}^6 \mid \sum_{i=1}^{6} x_i^2 = 1, x_1 x_6 - x_2 x_5 + x_3 x_4 = 0\} = \{x \in \mathbb{R}^6 \mid |x|^2 = 1, x \cdot \bar{x} = 0 \text{ where } \bar{x} = (x_6, -x_5, x_4, x_3, -x_2, x_1)\}$.

We can show that $M$ is a codimension 2 submanifold of $\mathbb{R}^6$. To prove this let $f: \mathbb{R}^6 \mapsto \mathbb{R}^2$ be defined by
$$f(x) = f(x_1, \ldots, x_6) = \left(\sum_{i=1}^{6} x_i^2, 2(x_1 x_6 - x_2 x_5 + x_3 x_4)\right) = (|x|^2, x \cdot \bar{x}), \text{ where}$$
$\bar{x} = (x_6, -x_5, x_4, x_3, -x_2, x_1)$. Then $M = f^{-1}(0,0)$ and we have
$$df_x = \begin{bmatrix} 2x_1 & 2x_2 & 2x_3 & 2x_4 & 2x_5 & 2x_6 \\ 2x_6 & -2x_5 & 2x_4 & 2x_3 & -2x_2 & 2x_1 \end{bmatrix}.$$
We show that the rank of $df_x$ on M is 2. Obviously $2 \geq rank(df_x) \geq 1$. If $rank(df_x) = 1$, Then we have
$$(x_1, \ldots, x_6) = \lambda(x_6, -x_5, x_4, x_3, -x_2, x_1) \Rightarrow x = \lambda \bar{x}, \qquad 0 \neq \lambda \in \mathbb{R},$$
Since on $M$, $x \cdot \bar{x} = 0$, then
$$x \cdot \lambda \bar{x} = \lambda |x|^2 = 0 \Rightarrow |x|^2 = 0.$$
But we know that $|x|^2 = 1$, so this is a contradiction. Thus $rank(df_x) = 2$ and $M$ is a submanifold of $\mathbb{R}^6$ with $\dim(M) = 6 - 2 = 4$.

Let $M$ be a differentiable manifold of class $C^\infty$ and with the dimension $m \geq 2$. If there exists on $M$ a tensor $J_i^j$ of type (1,1) satisfying
$$J_i^j J_j^k = -\varepsilon_i^k = \begin{cases} 1 & k = i, \\ 0 & k \neq i. \end{cases} \quad k, i, j = 1, \ldots, m, . \tag{4.1}$$
Then $J_i^j$ is said to define an *almost complex structure* on $M$ and manifold $M$ is called an almost complex manifold.

From (4.1) it follows that the almost complex structure $J_i^j$ induces an endomorphism $J: T_p M \mapsto T_p M$ of tangent space $T_p M$ of $M$ at each point $p$ of $M$ where $J^2 = -I$, $I$ being the identity operator on $T_p M$.

**Theorem 4.1.** There exists an almost complex structure on 4-dimensional submanifold
$$M = \{x \in \mathbb{R}^6 \mid |x|^2 = 1, x \cdot \bar{x} = 0\}.$$

**Proof.** If $p \in M$ then
$$T_p M = \{(v_1, \ldots, v_6) \in \mathbb{R}_p^6 \mid p \cdot v = 0, \quad p \cdot \bar{v} = 0\}$$
Because for each $v \in T_p M$ there exists a curve $\gamma$ on $M$ with $\gamma(0) = p$ and $\gamma'(0) = v$ but this main that
$|\gamma(t)|^2 = 1$ and $\gamma(t) \cdot \bar{\gamma}(t) = 0$ \hfill (4.2)
From (4.2) it follows that $\gamma(t) \cdot \gamma'(t) = 0$ and $\gamma(t) \cdot (\bar{\gamma}(t))' + \gamma'(t) \cdot \bar{\gamma}(t) = 0$,
But $\gamma(0) = p$ and $\gamma'(0) = v$, so $p \cdot v = 0$ and $2p \cdot \bar{v} = 0$.

Now let $J_p(v) = p \times v$ where $p \times v$ be defined by Definition 2.1. We have
$J_p^2(v) = J_p\left(J_p(v)\right) = J_p(p \times v) = p \times (p \times v) = -(p \times v) \times p =$
$-[(p \cdot p)v - (p \cdot v)p + (p \cdot \bar{p})v - (p \cdot \bar{v})\bar{p}]$.
Since $p \in M$, $v \in T_pM$ we have $p \cdot p = 1$ and $p \cdot \bar{p} = p \cdot v = p \cdot \bar{v} = 0$. Thus $J_p^2(v) = -(p \cdot p)v = -v$. So then $J_p^2 = -I$. ∎

In order to establish contact with the notation usually used in physics, we introduce in each admissible basis $\{e_\alpha\}$, the two 3-vectors $\vec{E} = E_1 e_1 + E_2 e_2 + E_3 e_3$ and $\vec{B} = B_1 e_1 + B_2 e_2 + B_3 e_3$, then we can define

$$F = [F_b^a] = \begin{bmatrix} 0 & B_3 & -B_2 & E_1 \\ -B_3 & 0 & B_1 & E_2 \\ B_2 & -B_1 & 0 & E_3 \\ E_1 & E_2 & E_3 & 0 \end{bmatrix}.$$

$F$ can be regarded as an electromagnetic field at some points of $\mathbb{R}^4$. In this case $\vec{E}, \vec{B}$ are the classical electric and magnetic fields 3-vectors at the point as measured in $\{e_\alpha\}$. Since $F \in So(4)$ we can related to $F$ the following matrix $A \in Spin(4)$ (or $A \in cl(1,3)$):

$$\begin{bmatrix} E_0 + iE_1 & B_0 + iB_1 & B_3 + iE_2 & -B_2 + iE_3 \\ B_0 + iB_1 & E_0 + iE_1 & E_3 + iB_2 & -E_2 + iB_3 \\ -B_3 + iE_2 & -E_3 + iB_2 & E_0 - iE_1 & B_1 + iB_0 \\ B_2 + iE_3 & E_2 + iB_3 & -B_2 - iB_0 & E_0 - iE_1 \end{bmatrix}.$$

In the above representation we added two components $E_0$ and $B_0$.

**5. Conclusion.** We consider an outer product on a six dimensional real vector space, and we deduce Lie brackets on it to take the structures of so(4), so(2,2) and so(3,1). We also find an almost complex structure on a submanifold of it.